\documentstyle[sprocl,times,journals]{article}

\newcommand{\gsim}
	{\mathrel{\raise.25ex\hbox{$>$}\kern-.8em\lower.9ex\hbox{$\sim$}}}
\newcommand{\lsim}
	{\mathrel{\raise.25ex\hbox{$<$}\kern-.8em\lower.9ex\hbox{$\sim$}}}
\newcommand{\nearprop}
	{\mathrel{\raise.25ex\hbox{$\propto$}\kern-.8em\lower.9ex\hbox{$\sim$}}}
\newcommand{\const}{{\rm const}}

\newcommand{\muG}{{\,\mu\rm G}}

\newcommand{\EeV}{{\,\rm EeV}}
\newcommand{\ZeV}{{\,\rm ZeV}}

\newcommand{\pG}{{\,\rm pG}}

\newcommand{\Mpc}{{\,\rm Mpc}}

\begin{document}

\title{Propagation of UHE cosmic rays in a structured universe}
\author{J\"org P.~Rachen} 
\address{Pennsylvania State University, Astronomy Department\\
	525 Davey Lab, University Park, USA-PA 16802. jorg@astro.psu.edu}
\maketitle

\abstracts{In a gravitationally unstable universe, the structure of dark
matter and galaxies, intergalactic gas and magnetic field can have severe
impact on the propagation of ultra high energy cosmic rays
(UHECR).\cite{RK96} The possible effects include spatial confinement and
directional focusing along the supergalactic matter sheets, as well as
universal re-acceleration at large scale shock fronts, and spectral
modification due to energy dependent leakage into cosmic voids. As a result,
the GZK-cutoff may be less pronounced and occur at a higher energy, where the
stochastic nature of both acceleration and energy loss processes has to be
taken into account.}

\section{Supergalactic magnetic field structure and cosmic ray confinement}
\vspace{-4pt}

Very little is known about the strength and orientation of the magnetic field
outside our galaxy. For cosmic ray transport calculations, one mostly uses
the assumption of a nanogauss field, which is homogeneous over cells with
some reversal scale of order $1\Mpc$. In such fields, the highest energy
cosmic ray protons have a gyro-radius $r_{\rm g}\sim 300\Mpc$, thus propagate
almost in straight lines; this opens the possibility of an ``UHECR
astronomy'', as anticipated by the Pierre Auger Project.\cite{Horvath}

Models of structure formation in cosmology, however, draw a quite different
picture: The magnetic field is aligned with the matter sheets, where it can
reach a field strength up to ${\sim}\muG$, while in the large cosmic voids
the field drops to its primordial value of ${\lsim}\pG$;\,\cite{RK96} this
scenario is fully consistent with existing observations.\cite{Kronberg} In
the sheets, which have a typical thickness of ${\sim}10\Mpc$, the highest
energy cosmic rays have $r_{\rm g}\sim 1\Mpc$, and are thus confined. Outside
the sheets, the accretion flow of intergalactic gas drives the cosmic rays
back, but the rapidly decreasing magnetic field may allow diffusive losses in
upstream direction, which can imply spectral modifications due to a ``leaky
box'' mechanism. Fringe field effects may additionally focus and align the
cosmic rays with the field direction in the sheets; this might explain the
apparent correlation of UHECR arrival directions with the local sheet, the
``supergalactic plane''.\cite{SBLRW95,Hay96} Since the universe needs no
longer to be homogeneously filled with cosmic rays, the total energy budget
for UHECR sources is strongly diminished.

\vspace{-4pt}
\section{Large scale shocks and universal acceleration}
\vspace{-4pt}

Another prediction in a structured universe is the existence of large scale
shock fronts,\cite{RK96} providing the possibility of cosmic ray acceleration
by the very effective shock-drift acceleration mechanism.\cite{Jok87} In a
global picture, the matter sheets form the collective downstream region, and
the voids the collective upstream region in a foam-like shock topology. The
cosmic rays, sliding sideways along the shock, never effectively leave the
acceleration region. The spatial extension of the acceleration region in the
direction of the flow can be estimated by the diffusion length, $l_{\rm D}$,
which depends on particle energy for quasi-perpendicular shocks and a
Kolmogoroff turbulence spectrum as $l_{\em D} \nearprop E^{5/3}$. At the
highest energies, $l_{\rm D}$ can become comparable to the sheet thickness,
and the particles scatter freely between the boundary shock fronts. In this
case, a stationary particle spectrum will no longer be obtained by the
balance of diffusion over the shock front and downstream advection, but
rather by the balance of energy gains and losses due to MBR
interactions; here, the stochastic nature of the loss process turns out to be
important.

\vspace{-4pt}
\section{The stochastic nature of MBR pion production losses}
\vspace{-4pt}

The transport of a proton in the MBR which is subject to pion production
losses has to be described by a Markov point process, where the energy loss
occurs randomly in distinct steps of random-distributed width. We may
simplify the process to a pure counting process of unit steps, which is in
case of a constant interaction rate known as a Poisson process. In photopion
interactions, particles lose energy fractionally, i.e $\Delta E/E
\,{\approx}\,\Delta{\ln}E\,{=}\,\const\,{\approx}\,0.2$. For a Poisson
process, one can show that an initial spectrum power law spectrum, $f\propto
E^{-a}$, of a source at distance $D$, suffers an energy independent reduction
by a factor $M = \exp[-(D/\lambda) (1 - e^{-\alpha})]$, if $\lambda$ is the
mean interaction length and $\alpha = a(\Delta{\ln}E)$.  For a linearly
increasing interaction rate, $\rho = c/\lambda = \rho'\ln (E/E_0)$, the
modified spectrum can be approximately described as a power law steepened by
$\Delta a = (D\rho'/c) (1 - e^{-\alpha})$.\cite{PhD}

The interaction rate in the microwave background can be best modeled relative
to the maximum rate $\rho_1$, which is reached for $E>E_1\,{\approx}\,1\ZeV$
and corresponds to $\lambda_1\,{\approx}\,4\Mpc$. For
$E_0\,{\approx}\,30\EeV<E<E_1$ it is linearly increasing,
$(\rho/\rho_1)\approx 0.3 \ln(E/E_0)$, and $\rho=0$ for $E<E_0$. A continuing
initial power law spectrum maps then to a piecewise power law with index $a$
for $E<E_0$ and $E>E_1$, and $a +
\Delta a$ in between. A spectral cutoff in the source maps to an exponential
decline of the observed spectrum somewhat below the source cutoff
energy.\cite{PhD} We may give two numerical examples: A radio galaxy at
$D=30\Mpc$, producing a spectrum $f\propto E^{-2}$ with a sharp cutoff at
$1\ZeV$, is observed with a power law index $a'{\approx}2.75$ between $30$
and $300\EeV$, followed by an exponential cutoff. A topological defect at
$D=100\Mpc$, producing a $f\propto E^{-1.3}$ spectrum, is observed with a
power law index $a'{\approx}3$ between $30\EeV$ and $1\ZeV$, flattening back
to $a{=}2$ for higher energies.

\vspace{-4pt}
\section{Consequences for the GZK cutoff and cosmic ray observatories}
\vspace{-4pt}

The time scale of large scale shock acceleration, $t_{\rm a}$, is generally
larger than the time scale for MBR photopion losses, $t_\pi$; depending
on magnetic field strength and shock\linebreak

\noindent velocity, we may find ratios $t_{\rm
a}/t_\pi\sim 1{-}100$ at ${\gsim}100\EeV$.\cite{KRB97} Thus the
acceleration is not really effective in the ordinary sense; however,
considering the stochastic nature of energy losses and the breakdown of
advection at the highest energies, the resulting stationary spectrum can
still be relatively flat for $E{\gsim}100\EeV$: In the simple case of a
constant acceleration time scale, interaction loss balanced shock
acceleration leads to power laws $f\propto E^{-b}$, and the relation $t_{\rm
a}/t_\pi = b[1 - \exp(-0.2 b)]^{-1}$ holds. Spectral indices as observed
in the UHECR spectrum are obtained for $t_{\rm a}/t_\pi\approx 4$, but
steepen very fast for larger values; under realistic conditions, the
equilibrium spectrum is probably concave and too steep to explain the highest
energy event rates.

Therefore, the existence of large scale shocks in the universe does not make
cosmic ray point sources unnecessary; radio galaxies, AGN, gamma ray bursters
or topological defects may still contribute as UHECR sources. Clusters of
galaxies, which are the sites of the strongest large scale shocks and well
located in the universe, can play an intermediate role between point sources
and large scale acceleration.\cite{KRB97} The importance of large scale
shocks is rather that they provide a {\em re-acceleration mechanism} which is
{\em as universal as the GZK process}, and thus may lead to revised estimates
of the maximum distance of the possible sources of highest energy cosmic
rays. Consequently, the pros and cons for the various source models have to
be reconsidered in a structured universe.  For the Pierre Auger UHECR
observatory, the large values of the magnetic field arising from large scale
structure simulations give little hope to see point sources of charged cosmic
ray particles; however, UHECR events {\sl are} likely to occur in clusters
and map the local large scale structure of the magnetic field.

\begin{small}
\section*{Acknowledgements}
\vspace{-3pt}
Work of JPR is funded by NASA grant 5-2857. This work is based on a PhD
thesis supervised by P.L.~Biermann at the MPIfR Bonn, and a collaboration
with H.~Kang. T.~Stanev is acknowledged for discussions.

\section*{References}
\vspace{-3pt}
\bibliographystyle{unsrt}
\bibliography{references}
\end{small}

\end{document}